\documentclass[aps,prl,notitlepage,superscriptaddress,showpacs,twocolumn ]{revtex4}
\usepackage{graphicx,subfigure,epsfig}
\usepackage{dcolumn}
\usepackage{amssymb}
\usepackage{times}
\usepackage{amsmath}
\usepackage{amsfonts}
\usepackage{mathrsfs}
\usepackage{setspace}
\usepackage{latexsym}
\usepackage{bbm}
\usepackage{float}
\usepackage{flafter}
\usepackage{bm}
\usepackage{epstopdf}
\usepackage{hyperref}
\usepackage{color}
\usepackage{multirow}

\begin{document}

\title{Numerical Study of Quantum Hall Bilayers at Total Filling $\nu_T=1$: \\
A New Phase at Intermediate Layer Distances}

\author{Zheng Zhu}
\affiliation{Department of Physics, Massachusetts Institute of Technology, Cambridge, MA, 02139, USA}
\author{Liang Fu}
\affiliation{Department of Physics, Massachusetts Institute of Technology, Cambridge, MA, 02139, USA}
\author{D. N. Sheng}
\affiliation{Department of Physics and Astronomy, California State University, Northridge, CA, 91330, USA}

\begin{abstract}
We study the phase diagram of quantum Hall bilayer systems with total filing $\nu_T=1/2+1/2$ of the lowest Landau level as a function of layer distances $d$.
Based on  numerical exact diagonalization calculations, we obtain three distinct phases,  including an exciton superfluid phase with spontaneous interlayer coherence at small $d$,  a composite Fermi liquid at large $d$, and an intermediate phase for $1.1<d/l_B<1.8$ ($l_B$ is the magnetic length).
The transition from the exciton superfluid to the intermediate phase is identified by (i) a dramatic change in the Berry curvature of the ground state under twisted boundary conditions on the two layers; (ii) an energy level crossing of the first excited state. The transition from the intermediate phase to the composite Fermi liquid is identified by the vanishing of the exciton superfluid stiffness. Furthermore, from our finite-size study, the energy cost of transferring one electron between the layers shows an even-odd effect and possibly extrapolates to a  finite value in the thermodynamic limit, indicating the enhanced intralayer correlation. Our identification of an intermediate phase and its distinctive features shed new light on the theoretical  understanding of the quantum Hall bilayer system at total filling $\nu_T=1$.
\end{abstract}


\maketitle

\emph{Introduction.}---The multilayer  quantum Hall systems  demonstrate tremendously rich physics when tuning the interlayer interaction by changing layer distance $d$. One of the  prominent examples is the bilayer systems\cite{Girvin,Eisenstein2004,Narozhny2016,Eisenstein2014} at a total filling $\nu_T=1$ ($\nu=1/2$ in each layer) with negligible tunneling.  Experimentally, the bilayer systems can be realized in single wide quantum wells, double quantum wells or bilayer graphenes \cite{bilayer1,bilayer2,graphene1,graphene2,graphene3}. Theoretically, the quantum states in small and large $d$ limits have been well understood.
When the layer distance is small, the strong interlayer coulomb interaction drives the electron system into
a pseudospin (layer)  ferromagnetic long range order (FMLRO)  state with the spontaneous interlayer phase coherence and interlayer superfluidity \cite{Wen1992,Moon1995,Yang1996,Balents2001,Stern2001}.
The  FMLRO can also be described as an exciton condensation state as an electron in an orbit of one layer is always bound to a hole in another layer forming an exciton pair.
This excitonic superfluid state can be described by Haplerin ``111 state" wavefunction \cite{Halperin1983,Yoshioka1989}.  In the limit of infinite layer separation,  the bilayer system reduces to two decoupled composite Fermi liquids (CFL)\cite{Halperin1993,Rezayi1994,Kalmeyer 1992,Jain,Son2015}.

Several theoretical scenarios\cite{Cote1992,Bonesteel1996,Kim2001,Joglekar2001,Stern2002,Veillette2002,Simon2003,Wang2003, Doretto2006, Alicea2009,Cipri2014,Isobe2016,Sodemann2016,Potter2016} have been proposed for understanding the transition between  the exciton superfluid and CFL at intermediate layer distances. Due to its non-perturbative nature,  controlled analytical method for this problem is still lacking, and numerical techniques have been playing an important role.  Some numerical studies report a single phase transition, or a crossover, between the small and large distance regimes\cite{Schliemann2001,Shibata2006,Sheng2003}. Meanwhile,  an intermediate phase is found in ED and variational studies  \cite{Park2004,Moller2008,Moller2009,Milovanovic2015},  where the $p$-wave paired composite fermions state \cite{Moller2008,Moller2009} is proposed. Until now it remains controversial for the phase at intermediate distances.

On the experimental side, transport measurements indicate a transition between an exciton condensed interlayer coherent incompressible
quantum Hall effect   state and compressible liquid with varying the  layer  distance\cite{Kellogg2004,Wiersma2004,Murphy1994,Giudici2010}. At smaller layer distance, the total Hall conductance is quantized to $e^2/h$. A strong enhancement in the zero-bias interlayer tunneling conductance \cite{Spielman2000} and the vanishing of the Hall counterflow resistance \cite{Kellogg2004,Tutuc2004} provide evidence for interlayer coherence \cite{Eisenstein2014}. Above a critical distance $d\approx 1.6\sim2$ (in units of magnetic length $l_B$) which depends on the quantum well thickness, a compressible liquid state is found \cite{Eisenstein2014,Spielman2000,Tutuc2004,Kellogg2004,Wiersma2004,Murphy1994,Giudici2010,Luin2005,Finck2010}. However, the nature of the state at the intermediate distance is unsettled after numerous investigations\cite{Eisenstein2014}.

\begin{figure}[tbp]
\begin{center}
\includegraphics[width=0.45\textwidth]{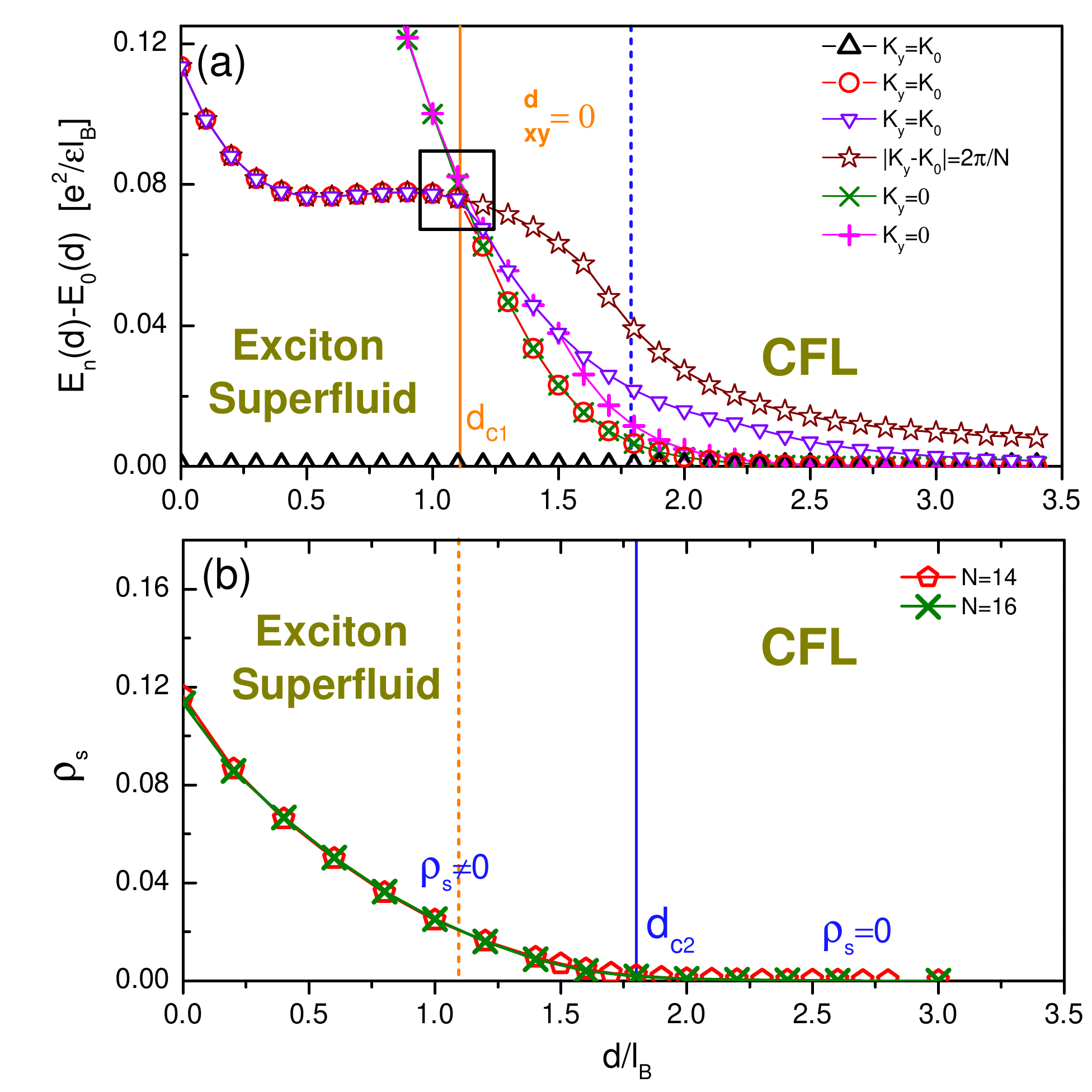}
\end{center}
\par
\renewcommand{\figurename}{Fig.}
\caption{(Color online) The phase diagram of $\nu=1/2+1/2$ quantum Hall bilayers with varying layer distance $d/l_B$. We identify three phases: exciton superfluid phase, the intermediate phase and composite Fermi liquid (CFL) phase. (a) The transition  from  exciton superfluid  to intermediate phase near $d_{c_1}\approx1.1$ is identified by the drag Hall conductance $\sigma^d_{xy}$ and the energy level crossing. Here, the ground state is  in the momentum sector $K_0=\pi$ and $N=16$ .   (b) The transition from intermediate phase to CFL phase near $d_{c_2}\approx1.8$ is identified by the exciton superfluid stiffness $\rho_s$ [see Eq.~\ref{Eq:rho_s}].}
\label{Fig:PhaseDia}
\end{figure}

Motivated by this unsolved issue,  we perform an extensive ED study of $\nu=1/2+1/2$ bilayer system on torus\cite{Rezayi2000, Haldane1985,Yoshioka84} up to 20 electrons,
the  phase diagram is summarized in Fig.~\ref{Fig:PhaseDia}.  We identify signatures of two phase transitions between the exciton superfluid  and the CFL at critical distances $d_{c_1}\approx 1.1$ and $d_{c_2}\approx 1.8$, respectively.
For layer distance $d<d_{c_1}$,  we establish  the exciton superfluid state by the existence of Goldstone mode, vanishing of single pesudospin excitation gap and finite exciton superfluid stiffness. Furthermore,  the Berry curvature shows strong fluctuation, leading to non-quantized drag Hall conductance which is consistent with the gapless feature. For the intermediate layer distance $d_{c_1}<d<d_{c_2}$,  we find the gapped single pseudospin excitation with even-odd effect, which is combined with a finite exciton superfluid stiffness. The drag Hall conductance is  quantized to  zero with no singularity in the Berry curvature, while the total Hall conductance remains exactly  quantized to $e^2/h$. The quantum phase transition between the exciton condensed  state  and intermediate phase is identified by a dramatic change in the Berry curvature of the ground state under twisted boundary conditions on the two layers, and
the level crossing with a  change of the nature of the low-lying excitations at $d=d_{c_1}$.  The fact of level crossing near $d_{c_1}$ is consistent with previous studies\cite{Park2004,Shibata2006,Milovanovic2015}. The second transition between the intermediate phase and the CFL is  characterized  by the  vanishing of the exciton superfluid stiffness. Further discussions of the finite size effect of numerical simulation can be found  in supplementary materials.

\begin{figure*}[tpb]
\begin{center}
\includegraphics[width=1\textwidth]{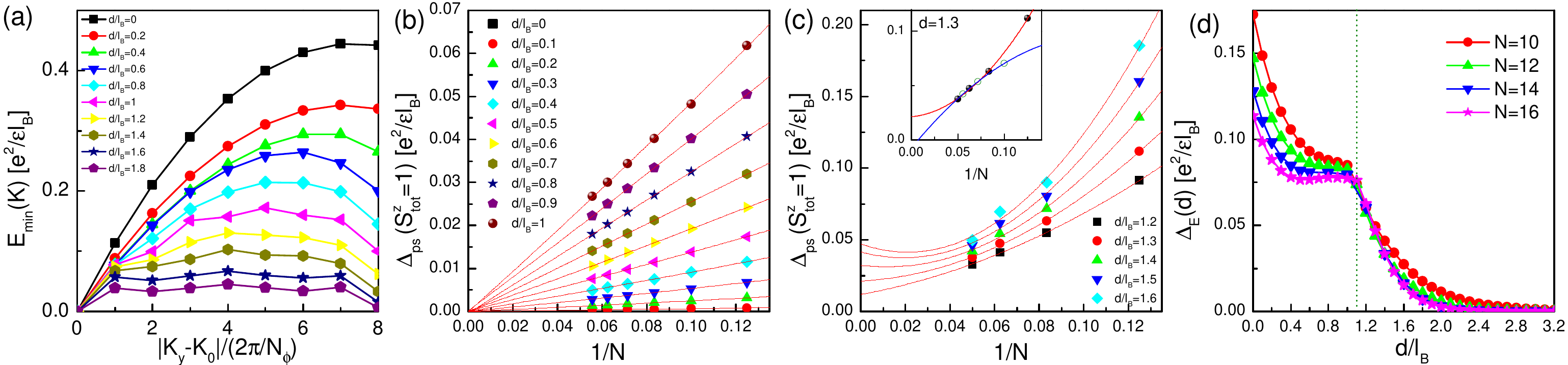}
\end{center}
\par
\renewcommand{\figurename}{Fig.}
\caption{(Color online) (a) The energy dispersion curves of lowest-energy excitations at each momentum sector.  Here, the ground state is  in the momentum sector $K_0=\pi$.   (b) and (c)  show finite size scaling of the single pseudospin excitation gap $\Delta_{ps}$ by using parabolic function for layer distance$d/l_B< 1.1$ (b) and $d/l_B> 1.1$ (c). The inset of (c) indicates the even-odd effect in the intermediate phase up to $N=20$. (d) The energy spectrum gap $\Delta_E\equiv E_1(d)-E_0(d)$ as a function of $d/l_B$. The cusp near $d/l_B\approx1.1$ indicates the level crossing for the excited states.}
\label{Fig:EK}
\end{figure*}

\emph{Model and Method.}--- We consider  bilayer electron systems subject to a magnetic field perpendicular to the two dimensional (2D) planes. We use torus geometry with the length vectors $\bf{L_x}$ and $\bf{L_y}$, and an  aspect angle $\theta$ between them. Here,  $L_x=L_y=L$ and $\theta=\pi/2$ for most of calculations. The magnetic length $l_B\equiv\sqrt{\hbar c/eB}\equiv1$ is set to be  the unit of the length and $N_\phi$ represents the number of magnetic flux quanta determined by $|L_xL_ysin\theta|=2\pi N_\phi$. In the presence of strong magnetic field, the Coulomb interaction, projected onto the lowest Landau level, is written as
 \begin{equation}\label {Ham}
V =\frac{1}{2\pi  N_\phi} \sum\limits_{i < j,\alpha ,\beta } {\sum\limits_{{\bf{q}},{\bf{q}} \ne 0} {{V_{\alpha \beta }}\left( q \right)} } {e^{ - {q^2}/2}}{e^{i{\bf{q}} \cdot \left( {{{\bf{R}}_{\alpha ,i}} - {{\bf{R}}_{\beta ,j}}} \right)}}.
  \end{equation}
Here, $\alpha (\beta)=1,2$ are indices of two layers (which are the  two components of a pseudospin 1/2), $V_{\alpha, \alpha}(q)= 2\pi {e^2}/({\varepsilon q})$ and $V_{12}(q)=V_{21}(q)= 2\pi {e^2} /({\varepsilon q})\cdot e^{-qd}$  are the Fourier transformations of the intralayer and interlayer Coulomb interactions, respectively. $d$ is the distance between two layers and  $\bf{R}_{\alpha,i}$ is the  guiding center coordinate of the $i$th electron in layer $\alpha$. In the present work, we  consider the physical systems with two identical 2D layers (with zero width) in the absence of  electron interlayer tunneling while  spins of electrons are fully polarized  due to strongly magnetic field.

We use ED algorithm to study the energy spectrum and state information on torus. In order to study the physics of the pseudospin sector, we generalize the periodical boundary condition to twisted boundary condition with  phase $0\leq\theta _\lambda ^\alpha\leq2\pi $ along $\lambda$ direction in the layer $\alpha$.
By a unitary transformation, one can get the the periodic wave function $\Psi $ on torus with
$|\Psi \rangle = \mathbf{exp} \left[ { - i\sum_\alpha  {\sum_i {\left( {(\theta _x^\alpha/L_x) x_i^\alpha  + (\theta _y^\alpha/L_y)y_i^\alpha } \right)} } } \right]|\Phi\rangle$.
Then the Berry curvature is defined by $ F(\theta _x^\alpha,\theta _y^\beta) =\mathbf{Im}( \left\langle  \partial \Psi/\partial {\theta^{\alpha} _x}  |  \partial \Psi/\partial {\theta^{\beta} _y} \right\rangle  - \left\langle  \partial \Psi/\partial {\theta^\beta _y }|\partial \Psi/\partial {\theta^{\alpha} _x}\right\rangle)$.  The integral  over the boundary phase unit cell leads to the topological Chern number matrix $C_{\alpha,\beta}=1/2\pi\int d\theta _x^\alpha d\theta _y^\beta F(\theta _x^\alpha,\theta _y^\beta)$, which contains topological  information for  the bilayer quantum Hall state\cite{Thouless1982,Niu1985,Wen1991,Yang2001,Sheng2003,Sheng2006,Sheng2011}.  Numerically, applying common and opposite boundary phases on two layers, one can obtain the Hall conductances in the layer symmetric and antisymmetric channel, denoted by $C^c (e^2/h)$ and $C^s (e^2/h)$, respectively. The drag Hall conductance, defined by $\sigma^d_{xy}=(C^c-C^s)(e^2/{2h})=(C_{1,2}+C_{2,1})(e^2/{h})$ ,  can be obtained directly by  calculating $C_{1,2}$ (or $C_{2,1}$), corresponding to  twisting boundary phases along $x$ direction in one layer and along $y$ direction in another layer. One can also obtain the exciton superfluid stiffness when applying twisted boundary phases\cite{Sheng2003}.

\emph{Energy Spectrum and Pseudospin Excitation Gap.}---In Fig.~\ref{Fig:EK} (a), we  show the lowest energies  in each momentum sector for different layer distances $d$. For smaller layer separations $d\lesssim 1.1$, indeed we find the low energy excitation has  the form of  linear dispersing Goldstone mode  for small momenta\cite{Spielman01}.  One can also measure the pseudospin excitation gap directly, which represents the energy cost of  moving one electron from one layer to another layer and is defined as $ \Delta_{ps}(d)\equiv E_0(N_\uparrow, N_\downarrow,d)-E_0(N/2,N/2,d)+d\cdot S^2_z/N_\phi$. Here, $N_\uparrow=N/2+\Delta N$ and $N_\downarrow=N/2-\Delta N$ denote the number of electrons in two layers for  $S_z=\Delta N=1, 2, \cdots$ excitation. The energy shift $d\cdot S^2_z/N_\phi$ is the charge energy induced by the imbalance of electron number in two layers with total pseudospin $S_z$\cite{MacDonald1990}. As shown in Fig.~\ref{Fig:EK} (b), the finite size scaling of $\Delta_{ps}(d)$ for  $S_z=1$  goes  to  zero in the  thermodynamic limit for $d\lesssim 1.1$.

As for layer distance $d \gtrsim 1.1$, the low energy linear dispersion spectrum  moves up in energy  [see  Fig.~\ref{Fig:EK} (a)] with new lower energy excitations appearing at other momenta sectors  for $d\gtrsim 1.1$  as shown in  Fig.~\ref{Fig:PhaseDia} (a).
For the layer distance  $d\thickapprox 1.1$, the energy spectrum  shows the level crossing of the first excited states between the $K_y=\pi$ (or $K_y=0$)
and $|K_y-K_0|=2\pi/N$ sectors (see Fig. \ref{Fig:PhaseDia}(a)).
Although the ground state still locates in $K_y=K_0$ sector at $d\thickapprox 1.1$, the level crossing for the first excited state indicates the change of the low-lying energy spectrum for the bilayer systems. Here, level crossing also characterizes a phase transition based on the indications of  pseudospin gap.

\begin{figure*}[htbp]
\begin{center}
\includegraphics[width=0.9\textwidth]{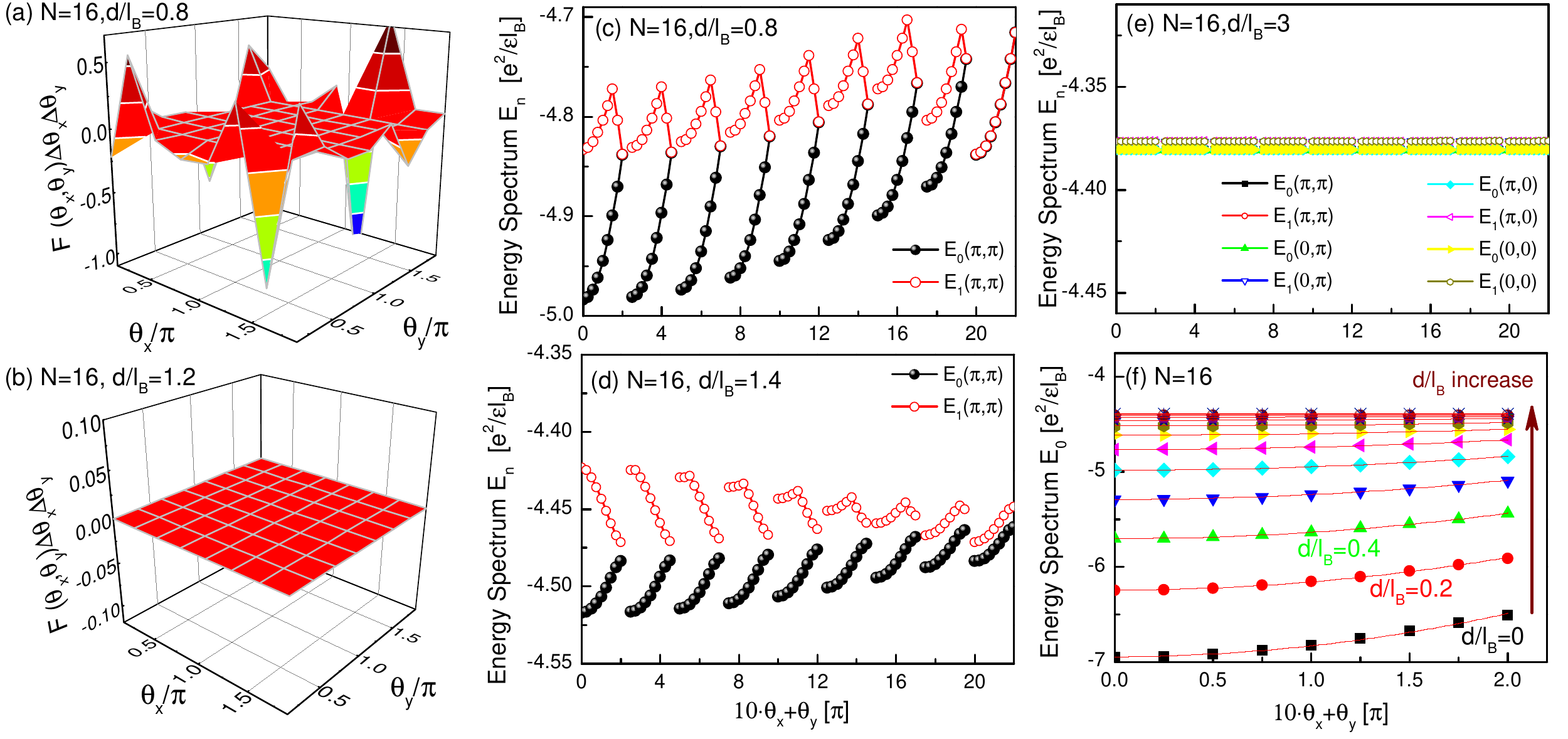}
\end{center}
\par
\renewcommand{\figurename}{Fig.}
\caption{(Color online) The Berry curvature $ F(\theta _x^\alpha,\theta _y^\beta)$ for $d/l_B$=0.8 (a) , $d/l_B$=1.2 (b). Here, $\Delta \theta_x$ and  $\Delta \theta_y$ are the interval of mesh in phase space. It has strong fluctuation in FMLRO phase (a), while it is smooth in the intermediate phase (b). (c) to (e) are  energy spectrum of $N=16$ system with twisted  boundary phases for  $d/l_B$=0.8 (c),  $d/l_B$=1.4 (d) and $d/l_B$=3 (e). By fitting the energy spectrum with twisted phases, one can get the exciton superfluid stiffness $\rho_s$  [see Eq.~\ref{Eq:rho_s}] (f), which decreases with the layer distance and finally vanishes for $d>d_{c_2}$. (f) From bottom to top, $d/l_B$ increases from $d/l_B=0$ with interval 0.2.}
\label{Fig:E_Twist}
\end{figure*}

For $d\gtrsim 1.1$, the $S_z=1$ pseudospin excitation displays even-odd effect determined by the electron number in each layer [see the inset of Fig.~\ref{Fig:EK} (c)],  indicating of the trend of intralayer pairing.
As shown in Fig.~\ref{Fig:EK} (c) with system sizes up to 20 electrons, the finite size scaling indicates gapped pesudospin excitation
for even electron number in each layer, while it is gapless when the electron number in each layer is odd. One should be careful in the fitting due to limited number of data points, however, the finite pseudo-spin excitation gap is also implied by the disappearance of linear dispersion mode [see  Fig.~\ref{Fig:EK} (a)] , the  flat Berry curvature, and well-defined spectrum gap when twisting boundary conditions [see Fig.~\ref{Fig:E_Twist}  (b) and (d) below].

Fig.~\ref{Fig:EK} (d) shows the energy gap $\Delta_E (d)\equiv E_1(d)-E_0(0)$ between two lowest energy states, one can find that the cusp due to the level crossing for the lowest  energy excitations near the transition point $d_{c_1}\thickapprox 1.1$ is  robust and independent on the lattice size, indicating the intrinsic property of such a transition. Clearly,  we have identified a transition from the gapless pseudospin  FMLRO state at smaller distance to the intermediate phase with new low-lying excitation and finite pseudospin gap.

\emph{Berry Curvature and Energy Spectrum Under Twisted Boundary Conditions.}---  The transition near $d_{c_1}\thickapprox 1.1$ can also be identified by the Berry curvature $ F(\theta _x^\alpha,\theta _y^\beta)$ and the energy spectrum under twisted boundary conditions. Physically, a gap state has a well-defined smooth Berry curvature, while a gapless state may have singular
Berry curvature associated with gapless points in low energy spectrum. Fig.~\ref{Fig:E_Twist} (a) and (b) show the Berry curvatures at the $d<d_{c_1}$ and $d_{c_1}<d<d_{c_2}$ by applying $\theta^1_x=\theta_x$, $\theta^2_x=0$ and $\theta^1_y=0$, $\theta^2_y=\theta_y$ for the lowest energy state in the sector $(\pi,\pi)$.  Fig.~\ref{Fig:E_Twist} (a) shows the strong fluctuation of the Berry curvature, suggesting the gapless pseudospin spectrum when $d<d_{c1}$.
The Berry phase is not well defined due to near level crossing (with Berry phase integrated over  each singular point
only defined up to the fractional part of $2\pi$), which  gives rise to the non-quantized drag Hall conductance in this regime\cite{Sheng2003}.  Since the Hall conductance in the symmetric channel is well defined in this regime, the non-quantized drag Hall conductance indicates gapless feature of the antisymmetric channel. On the other hand, the Berry curvature is near flat without any singularity in $d_{c_1}<d<d_{c_2}$ regime [see Fig. ~\ref{Fig:E_Twist} (b)], which is consistent with the well-defined single pseudospin excitation gap in this phase. Furthermore, the integral of the Berry curvature gives us zero drag Hall conductance in the intermediate phase, indicating the well defined Hall conductance in symmetric channel or finite charge gap in the intermediate phase.
We also find that  Berry curvatures in all four sectors  $(0,0)$ ,$(0,\pi)$,$(\pi,0)$,$(\pi,\pi)$  always have similar features
 and twisting boundary phases will connect   the ground state $(\pi,\pi)$ to the other three states.
In Fig.~\ref{Fig:E_Twist} (c) and (d), one can find the energy spectrum of the lowest two states in the same momentum sector $(K_x,K_y)=(\pi,\pi)$ with twisted phases. Here, we map the phase $\theta_x, \theta_y$ into one-dimensional quantity $\theta\equiv10\theta_x+\theta_y$ for convenience of plotting. The singularity in the Berry curvature for  $d<d_{c_1}$  origins from the energy level crossing as the bilayer relative boundary phase $\theta_y$ approaching $2\pi$ in contrast to the behavior in the $d>d_{c_1}$ regime, where a small gap opens to separate the lowest two states, indicating the existence of the pseudospin gap.
Based on the above analysis, we confirm that  the pseudospin Berry curvature also indicates the phase transition taking place near $d_{c_1}$.

\emph{Exciton Superfluid Stiffness.}---To study the evolution of  exciton superfluidity with the layer distances, we  obtain  the exciton superfluid stiffness $\rho_s$ by adding a small twisted boundary phase\cite{Sheng2003}, which is proportional to the superfluid density and identifies the energy cost when one rotates the order parameter of the magnetically ordered system by a small angle. In our ED calculation, the exciton superfluid stiffness can be obtained according to
 \begin{equation}\label{Eq:rho_s}
 E(\theta_t)/A=E(\theta_t=0)/A+\frac{1}{2}\rho_s \theta^2_t+O(\theta^4_t),
 \end{equation}
 where $E(\theta_t)$ is the ground-state energy with twisted
(opposite) boundary phases $\theta_t$ between two layers $\theta_t=\theta^1_x-\theta^2_x$ ($\theta^{1,2}_y=0$), $A=|{\bf L_x}\times{\bf L_y}|$ is the area of the torus surface. Fig.~\ref{Fig:E_Twist}  (c) to (e) show  the energy spectrum as a function of twisted phases for different layer distance. At smaller layer separation, one can find the ground state energy increases with tuning the twisted phases [see Fig.~\ref{Fig:E_Twist}  (c) and (d)]. By fitting the energy curve using the quadratic function [see Fig.~\ref{Fig:E_Twist}  (f)], we  get the exciton superfluid stiffness $\rho_s$, which decreases with the increase of the layer distance, and finally falls down to a negligible value for $d>d_{c_2}$ [see Fig.~\ref{Fig:PhaseDia} (b)].  As shown in Fig.~\ref{Fig:E_Twist}  (e), the energy almost does not change with the twisted phases for larger distances, indicating the vanish of superfluidity and the decoupling of two layers  for $d>d_{c_2}$, corresponding to CFL states.

\emph{Discussion.}--- We study the phase diagram of $\nu=1/2+1/2$ quantum Hall bilayers on torus and find that the exciton superfluid phase and CFL phase are separated by an intermediate phase, which exhibits finite exciton superfluid stiffness, flat Berry curvature,  zero drag Hall conductance and even-odd effect of pseudospins.

Theoretical interpretation of the intermediate phase may start from two well known limits. Starting from the infinite distance, it is nature to choose composite Fermion (CF) picture\cite{Isobe2016, Alicea2009,Cipri2014,Sodemann2016,You2017}. Recently, a fully gapped interlayer pairing phase is proposed based on random-phase approximation calculation\cite{Isobe2016}, which is consistent with our numerical findings of  flat Berry curvature as well as gapped spin-1 and charge excitations, but the explanation of finite exciton superfluid stiffness is lacking. The other candidate,  interlayer coherent CFL (ICCFL)\cite{Alicea2009} state, has  finite pseudospin stiffness due to  interlayer $U(1)$ phase fluctuations and possesses quantized Hall conductance in antisymmetric channel, which is consistent with our ED findings on finite pseudospin gap and flat Berry curvature. However, ICCFL indicates compressible property with respect to symmetric current, while our numerical data indicates finite charge gap as well as enhanced intralayer correlation [see supplementary material]. To understand the physics in charge channel better, one may start from the small distance limit in the composite boson (CB) picture\cite{Simon2003,Lian2017,Milovanovic2015} and assuming the system is $\nu=1$ integer quantum Hall state. Based on recent proposed wavefunction\cite{Lian2017}, the $SU(2)$ symmetry for CBs  emerges  near $d_{c_1}$, leading to the level crossing of first excited state[see Fig.\ref{Fig:PhaseDia}(a)]. The low-lying charge excitation is dominated by interlayer bound state of CB merons for $d<d_{c_1}$ while it is replaced by intra-layer bound state of CB merons for $d_{c_1}<d<d_{c_2}$, which explains the finite charge gap or quantized charge Hall conductance  and  the enhanced intra-layer correlations in the intermediate phase.

When taking both limits into account,  a mixed-state representation with considering both interlayer and intralayer correlations has been intensively studied\cite{Simon2003,Moller2008,Moller2009,Milovanovic2015,Milovanovic2017}. Such mixed representation leads to a $p$-wave interlayer pairing phase\cite{Moller2008,Moller2009} or the superfluid disordering phase \cite{Milovanovic2015} in the intermediate distance, which are consistent with  numerical finding of  the incompressibility in charge channel  and the disappearance of Goldstone mode  as the lowest energy excitation \cite{Milovanovic2017}. However, to explain all of numerical data consistently, it seems that one has to take into account the interplay between the  interlayer and intralayer correlations,
 which is still a theoretical challenge and calls for further theoretical study.

\begin{acknowledgments}
We acknowledge helpful discussions with I. Sodemann, T. Senthil, L.J. Zou, M. Zaletel, Z. Papi\'{c}, S. D. Geraedts, H. Isobe, Y.Z.You. Z.Z. and L.F. are supported by the David and Lucile Packard foundation.  L.F. is also supported by the DOE Office of Basic Energy Sciences, Division of Materials Sciences and Engineering under Award No. DE-SC0010526. D.N. Sheng is supported by the U.S. Department of Energy, Office of Basic Energy Sciences under grants No.  DE-FG02-06ER46305. D.N. Sheng was also supported in part by the Gordon and Betty Moore Foundation's EPiQS Initiative, grant GBMF4303 during her visit at MIT. Part of the simulation were preformed by using the Extreme Science and Engineering Discovery Environment (XSEDE), which is supported by National Science Foundation grant number ACI-1053575.
\end{acknowledgments}

\onecolumngrid

\begin{center}
{\bf {Numerical Study of Quantum Hall Bilayers at Total Filling $\nu_T=1$: A New Phase at Intermediate Layer Distances\\
Supplementary Material}}
\end{center}

\renewcommand{\theequation}{S\arabic{equation}}
\setcounter{equation}{0}
\renewcommand{\thefigure}{S\arabic{figure}}
\setcounter{figure}{0}
\renewcommand{\bibnumfmt}[1]{[S#1]}

\section{Exact Diagonalization Algorithm on Torus}
In this section, we introduce the application of exact diagonalization (ED) algorithm on the bilayer quantum Hall systems with torus geometry.  Considering $N$ electrons moving on the torus subject to a magnetic field perpendicular to its surface,  the length vectors of torus are ${\bf  L_x}$ and ${\bf L_y}$ with angle $\theta$ between them. We choose Landau gauge $ \mathbf{A}=(By,0,0)$ and have
\begin{equation}\label{eqS1}
 |L_xL_ysin\theta|=2\pi N_\phi.
\end{equation}
Here,  the magnetic length $l_B\equiv\sqrt{\hbar c/eB}\equiv1$ (the unit of length) and $N_\phi$ represents the number of magnetic flux quanta through the surface. We use the single-particle wave functions in the lowest Landau level (LLL) as basis, which reads
\begin{equation}
\psi_{k} (x,y) =\frac 1{\sqrt{L_x \sqrt{\pi}}}\sum_{n=-\infty}^{+\infty}e^{[\mathrm{i}(k+nL_y)x-(y+nL_y+k)^{2}/2]},
\label{wave}
\end{equation}
where $k\equiv  {2\pi j}/L_x $ with $j=0,1,...,N_\phi-1$ due to periodical boundary condition along $x$ direction. The single particle states $\psi_{k}$ are centered at $y=-k$ with a distance $2\pi/L_x$ apart along $y$ direction, while they are extended in $x$ direction. Then $N_\phi$ states can be mapped into one-dimensional (1D) lattice with each site representing a single particle orbital $\psi_{k}$. Then one can perform numerical simulation on such 1D lattice in momentum space with the number of sites equals to the number of orbitals. The relationship of the area of torus and the size of 1D lattice is determined by Eq.\ref{eqS1}.

In order to realize the numerical diagonalization on larger system size, one needs to reduce the dimension of the Hamiltonian block by taking advantage of magnetic translational symmetries along $x$ or/and $y$ directions. The symmetry analysis  was first provided by Haldane \cite{Haldane1985} with introducing two translation operators,  $T_\alpha$ $(\alpha =1,2)$ with eigenvalues $e^{2\pi iK_{\lambda} /N_\phi}$ ($ \lambda=x,y$  and $ K_{\lambda}=0,...,N_\phi-1$) . $T_1$ corresponds to the magnetic translation in $x$-direction, where $K_x=\sum_ {k=0}^{N_\phi-1} k n_k$ (mod $N_\phi$)  is total momentum (in the unit of $2\pi /L_x$) of electrons taken modulo $N_\phi$.  $T_2$ translates the entire lattice configuration one step $L_y/N_\phi=2\pi/L_x$ to the right along $y$-direction. For filling factor $\nu=N_e/N_\phi=p/q$ ($p$ and $q$ are coprime numbers), center of mass translations $T_2^s$ ($s=0,1, \dots q-1$) generate $q$-fold degenerate states. Then the energy eigenstates can be labeled by a two-dimensional vector $K_\lambda=0,...,N_\phi/q-1$. Taking advantages of one or both symmetries, one can numerically diagonalize the Hamiltonian efficiently. Different from the sphere geometry, there is no orbital number shift on torus and the states are uniquely determined by their filling factor.

For the bilayer systems discussed in this paper, we neglect the interlayer tunneling and only consider the models with intralayer and interlayer Coulomb interactions. By projecting the Coulomb interaction into the lowest Landau levels (LLL), we have the Hamiltonian
 \begin{equation}\label {Ham}
V =\frac{1}{2\pi  N_\phi} \sum\limits_{i < j,\alpha ,\beta } {\sum\limits_{{\bf{q}},{\bf{q}} \ne 0} {{V_{\alpha \beta }}\left( q \right)} } {e^{ - {q^2}/2}}  L^2_{n=0}[- {q^2}/2]{e^{i{\bf{q}} \cdot \left( {{{\bf{R}}_{\alpha ,i}} - {{\bf{R}}_{\beta ,j}}} \right)}}.
  \end{equation}
Here,  $\alpha (\beta)=1,2$ are indices of two layers (which are the  two components of a pseudospin 1/2) and $\bf{R}_{\alpha,i}$ is the  guiding center coordinate of the $i$th electron in layer $\alpha$, $L_n(x)$ is the Laguerre polynomial. $V_{\alpha, \alpha}(q)= 2\pi {e^2}/({\varepsilon q})$ and $V_{12}(q)=V_{21}(q)= 2\pi {e^2} /({\varepsilon q})\cdot e^{-qd}$  are the Fourier transformations of the intralayer and interlayer Coulomb interactions, respectively. $d$ is the distance between two layers. Numerically, one needs to use the second-quantization form:
 \begin{equation}
 V = \sum\limits_{{j_1}{j_2}{j_3}{j_4}} {{V^{\alpha \beta }_{{j_1}{j_2}{j_3}{j_4}}}c_{{j_1}}^\dag c_{{j_2}}^\dag {c_{{j_3}}}{c_{{j_4}}}},
 \end{equation}
 with
 \begin{equation}\label{ED}
 {V^{\alpha \beta }_{{j_1}{j_2}{j_3}{j_4}}} = {{\delta '}_{{j_1} + {j_2},{j_3} + {j_4}}}\frac{1}{{4\pi  N_\phi}}\sum\limits_{{\bf{q}},{\bf{q}} \ne 0} {{{\delta '}_{{j_1} - {j_4},{{{q_y}{L_y}} \mathord{\left/
 {\vphantom {{{q_y}{L_y}} {2\pi }}} \right.
 \kern-\nulldelimiterspace} {2\pi }}}}} V_{\alpha \beta}(q)\exp \left[ {{{ - {q^2}} \mathord{\left/
 {\vphantom {{ - {q^2}} 2}} \right.
 \kern-\nulldelimiterspace} 2} - i\left( {{j_1} - {j_3}} \right){{{q_x}{L_x}} \mathord{\left/
 {\vphantom {{{q_x}{L_x}} {{N_\phi }}}} \right.
 \kern-\nulldelimiterspace} {{N_\phi }}}} \right] L^2_{n=0}[- {q^2}/2].
     \end{equation}
Here, the Kronecker delta with the prime means that the equation is defined modulo $N_\phi$.  We also consider a uniform and positive background charge so that the Coulomb interaction at $q=0$ are canceled out\cite{Yoshioka84}.  Then the task of ED is to diagonalize the Hamiltonian in Eq. \ref{ED}.

\section{Twisting boundary conditions and the Hall Conductance}\label{Chern}

In this section, we introduce the numerical realization of calculating Chern number ($C$) or Hall conductance ($C\cdot e^2/h$) by twisting boundary conditions. For bilayer quantum Hall systems, twisting boundary conditions leads to $2\times 2$ Chern number matrix. The original treatment of Hall conductance as a topological invariant was firstly proposed in noninteracting electrons in a periodic potential\cite{Thouless1982}, and later it was generalized to interacting systems without periodicity\cite{Niu1985,Sheng2003,Sheng2006}. Followed the previous work\cite{Sheng2003,Sheng2006}, we begin with generalizing the periodical boundary condition to twisted boundary condition $T^\alpha (\bf{L}_\lambda)\left| \Phi  \right\rangle  = {e^{i\theta _\lambda ^\alpha }}\left| \Phi  \right\rangle$ for each electron , where  ${T^\alpha }(\bf{L}_\lambda)$  is the magnetic translational operator and $\lambda= x,y$. The twisted phases $0\leq\theta _\lambda ^\alpha\leq2\pi $ is along $\lambda$ direction in the layer $\alpha$. By a unitary transformation, one can get the the periodic wave function $\Psi $ on torus with
 \begin{equation}
|\Psi \rangle = \exp \left[ { - i\sum_\alpha  {\sum_i {\left( {\frac{\theta _x^\alpha}{L_x} x_i^\alpha  + \frac{\theta _y^\alpha}{L_y}y_i^\alpha } \right)} } } \right]|\Phi\rangle.
 \end{equation}
 Then the Berry curvature is defined by
 \begin{align}
  F(\theta _x^\alpha,\theta _y^\beta) =\mathbf{Im}\left(\langle{\frac{\partial\Psi}{\partial\theta^{\alpha}_x}}|{\frac{\partial\Psi}{\partial\theta^{\beta}_y}}\rangle
-\langle{\frac{\partial\Psi}{\partial\theta^{\beta}_y}}|{\frac{\partial\Psi}{\partial\theta^{\alpha}_x}}\rangle\right).
\end{align}
The integral  over the boundary phase unit cell leads to the topological Chern number matrix
 \begin{equation}
C_{\alpha,\beta}=\frac{1}{2\pi}\int d\theta^{\alpha}_{x}d\theta^{\beta}_{y} F(\theta _x^\alpha,\theta _y^\beta).
   \end{equation}

Numerically, applying common and opposite boundary phases on two layers, one can obtain the Hall conductances in the layer symmetricand antisymmetric channel, denoted by $C^c (e^2/h)$ (charge Hall conductance) and $C^s (e^2/h)$ (spin Hall conductance), respectively. In particular, with $\theta^1_{x}=\theta^2_{x}=\theta_{x}$ and $\theta^1_{y}=\theta^2_{y}=\theta_{y}$, the Chern number corresponds to the Hall conductances in the layer symmetric channel or the so-called charge Hall conductance,
\begin{align}
  C^{c}&=\int\frac{d\theta_{x}d\theta_{y}}{2\pi}\mathbf{Im}
  \left(\langle{\frac{\partial\Psi}{\partial\theta_x}}|{\frac{\partial\Psi}{\partial\theta_y}}\rangle
  -\langle{\frac{\partial\Psi}{\partial\theta_y}}|{\frac{\partial\Psi}{\partial\theta_x}}\rangle\right)\nonumber\\
& =C_{11}+C_{12}+C_{21}+C_{22}=\nu.
\end{align}
With $\theta^1_{x}=-\theta^2_{x}=\theta_{x}$ and $\theta^1_{y}=-\theta^2_{y}=\theta_{y}$, the Chern number corresponds to  the Hall conductances in the layer antisymmetric channel or the so-called the spin Hall conductance,
\begin{align}
  C^{s}&=\int\frac{d\theta^{x}d\theta^{y}}{2\pi}\mathbf{Im}
  \left(\langle{\frac{\partial\Psi}{\partial\theta_x}}|{\frac{\partial\Psi}{\partial\theta_y}}\rangle
  -\langle{\frac{\partial\Psi}{\partial\theta_y}}|{\frac{\partial\Psi}{\partial\theta_x}}\rangle\right)\nonumber\\
& =C_{11}-C_{12}-C_{21}+C_{22}.
\end{align}
The drag Hall conductance, defined by
 \begin{equation}
\sigma^d_{xy}=(C^c-C^s)\frac {e^2}{2h}=(C_{1,2}+C_{2,1})\frac {e^2}{h},
  \end{equation}
 can be obtained directly by  calculating $C_{1,2}$ (or $C_{2,1}$), corresponding to  twisting boundary phases along $x$ direction in one layer and along $y$ direction in another layer.

\section{Charge Gap and intralayer pairing}

For the intermediate phase we identified in the main text, one typical feature is that the drag Hall conductance is zero. As shown in Sec.\ref{Chern}, the drag Hall conductance equals to the Hall conductance in the symmetric (charge) channel minus the Hall conductance in the antisymmetric (spin) channel, the well defined zero drag Hall conductance indicates the quantum Hall conductance in the symmetric and antisymmetric channels are cancelled with each other.

\begin{figure}[htbp]
\begin{center}
\includegraphics[width=0.4\textwidth]{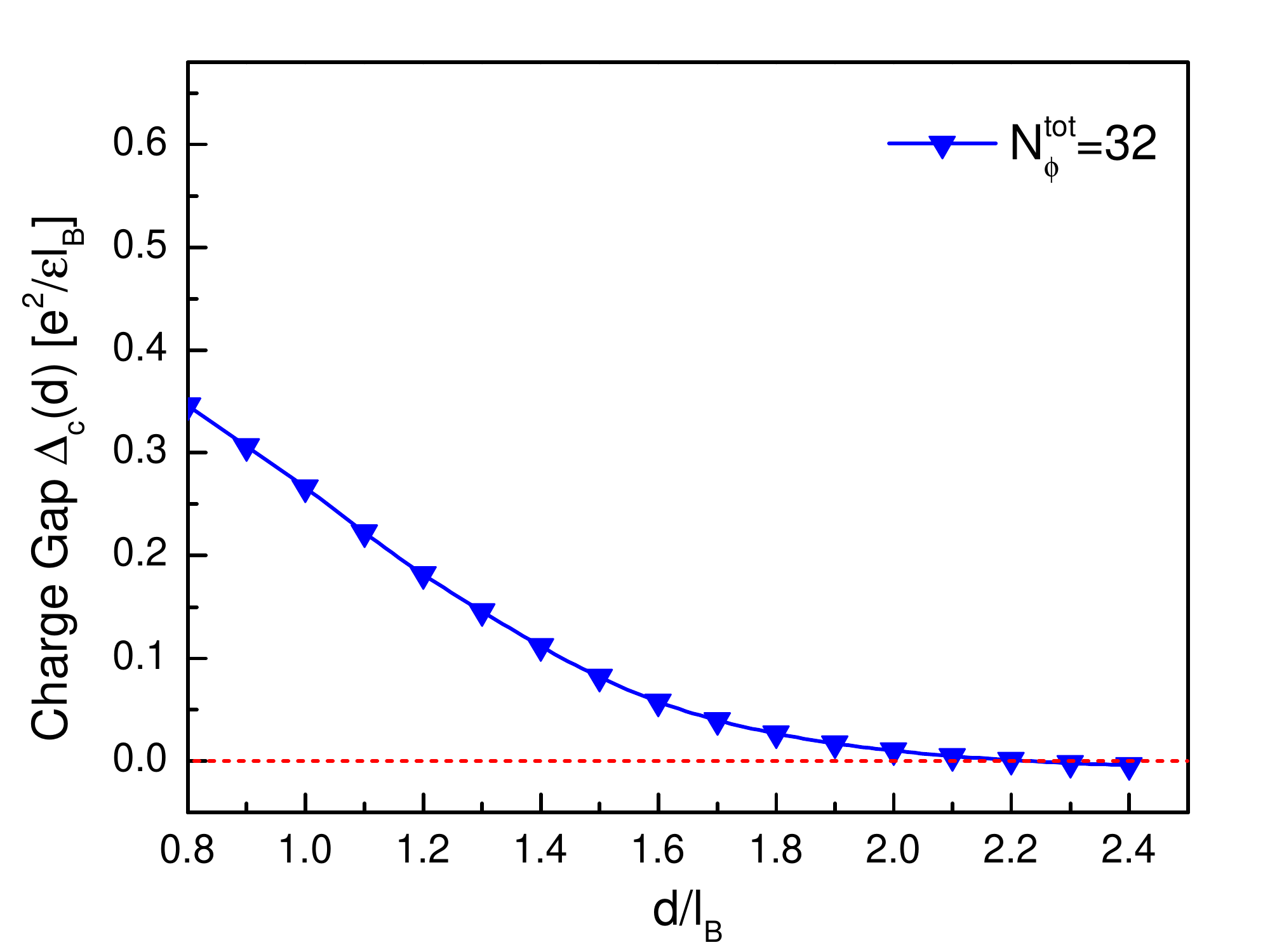}
\end{center}
\par
\renewcommand{\figurename}{Fig.}
\caption{(Color online) The charge gap as a function of  layer distance $d/l_B$ for $N=16$ electrons.}
\label{Fig:ChargeGap}
\end{figure}

As we have shown in Fig. 2  in the intermediate phase,  the finite pseudospin gap  is proved  from the consistency of the following aspects: (i) the finite size scaling analysis of the pseudospin gap [see Fig.2(c)]; (ii) the flat feature in the Berry curvature [see Fig.3(b)]; (iii) the robust gap between the lowest energy sates in the energy flow diagram when twisting boundary conditions [see Fig.3(d)]. Then the quantum Hall conductance in antisymmetric channel is well defined, leading to the well defined gap for charge excitations due to zero drag Hall conductance. To prove such indication explicitly, we directly measure the charge gap, which is defined by $ \Delta_{c}(d)\equiv E_0(N_\uparrow+1, N_\downarrow+1,d)+E_0(N_\uparrow-1, N_\downarrow-1,d)-2E_0(N_\uparrow,N_\downarrow,d)$. Here, $N_\uparrow$ and $N_\downarrow$ denote the number of electrons in two layers. As shown in Fig.\ref{Fig:ChargeGap}, both the ``111 state" and intermediate phase has finite gap for charge excitations.

\begin{figure}[htbp]
\begin{center}
\includegraphics[width=0.7\textwidth]{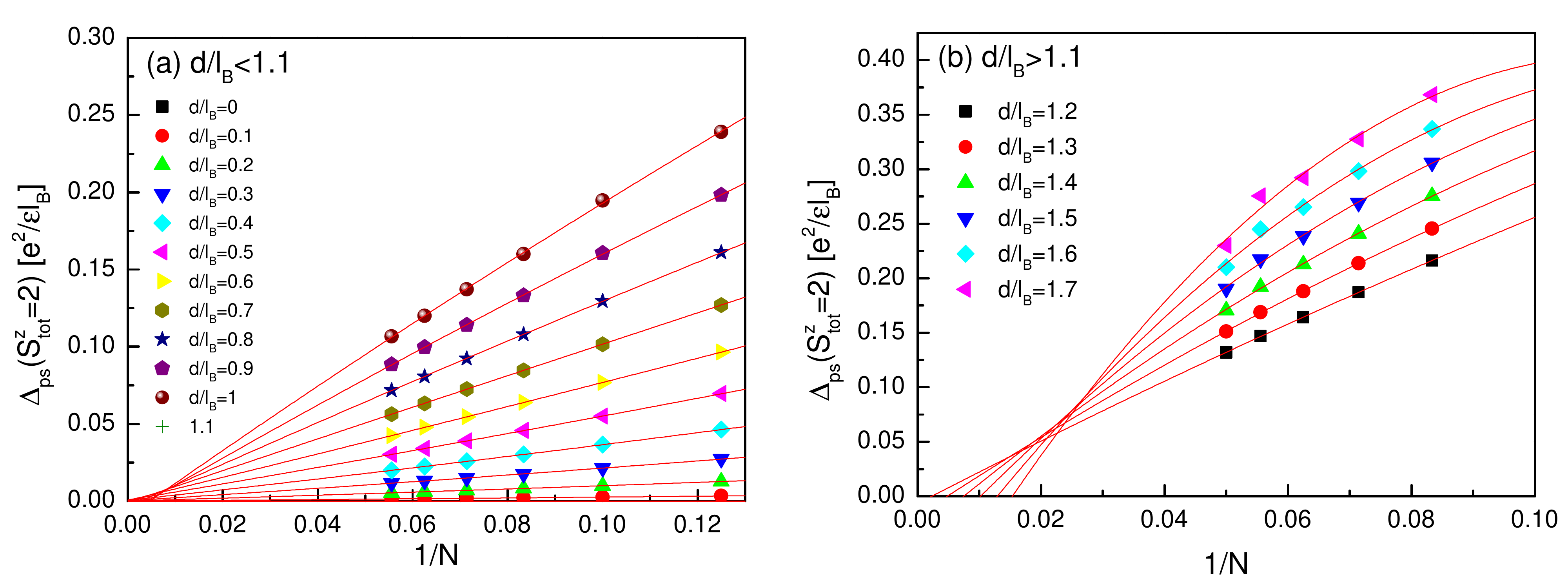}
\end{center}
\par
\renewcommand{\figurename}{Fig.}
\caption{(Color online) Finite size scaling of the two pseudospins excitation gap $\Delta_{ps}$ by using parabolic function for layer distance $d/l_B< 1.1$ (a) and $d/l_B> 1.1$ (b).}
\label{Fig:Sz=2}
\end{figure}

Besides the charge gap,  the energy cost of transferring one electron between the layers displays an even-odd effect, which is an indication of intra-layer pairing or enhanced intralayer correlations. For the case with even number of electrons in each layer, the gap is finite because moving one electron from one layer to the other will cost energy to break the pair, while such energy cost vanishes for the case with odd number of electrons in each layer. To check this property explicitly, we directly measure the pseudo-spin $S_z=2$ gap, which corresponds to move two electrons from one layer to the other. Numerically, the pseudospin gap is defined as $ \Delta_{ps}(d)\equiv E_0(N_\uparrow, N_\downarrow,d)-E_0(N/2,N/2,d)+d\cdot S^2_z/N_\phi$.  Here, $N_\uparrow=N/2+\Delta N$ and $N_\downarrow=N/2-\Delta N$ denote the number of electrons in two layers for  $S_z=\Delta N=1, 2, \cdots$ excitation. The energy shift $d\cdot S^2_z/N_\phi$ is the charge energy induced by the imbalance of electron number in two layers with total pseudospin $S_z$\cite{MacDonald1990}. As shown in Fig.\ref{Fig:Sz=2}, For the exciton superfluid phase at $d/l_B<1.1$ regime [see Fig.\ref{Fig:Sz=2} (a)], the finite size scaling also indicates the vanishing of the spin $S_z=2$ gap, which is similar to flipping single pseudospin, indicating the gapless property in spin channel for ``111 state".  However, for the intermediate phase at $1.1<d/l_B<1.8$ regime[see Fig.\ref{Fig:Sz=2} (b)], the even-odd effect disappears and the finite size scaling indicates the spin gap for $S_z=2$ is zero. Based on the above two aspects, the existence of intra-layer pairing state is indicated numerically.

Here, we add some comments on earlier study of the intermediate phase in this system.  Earlier ED study of such systems on sphere shows that the excitonic superfluid phase and the CFL phase  are possibly connected via an intermediate phase characterized by the $p$-wave paired composite fermions \cite{Moller2008,Moller2009}, Such interlayer $p$-wave paired states has finite charge gap, which is consistent with numerical result, however, we also found the features such as the coexistence of finite pseudospin gap and intralayer pairing state, which has not been discussed  in this framework. In an earlier  ED simulation on torus\cite{Sheng2003} with including random disorder scattering (limited to electron number $N=12$), a phase diagram with a direct transition from the superfluid phase to   the CFL was obtained,  which may be consistent with current results as the superfluid density is indeed nonzero up to the critical point $d_{c_2}$ where a transition to the CFL takes place. The weak disorder system may have two scenarios, one is that  the  pseudospin gap may be nonzero for the intermediate regime indicating two phases inside the superfluid regime, or the gapped  phase may have a transition back to a weak superfluid phase with gapless excitations driven by weak disorder scattering  as the pseudospin gap is very small even in the pure limit.

\begin{figure*}[htbp]
\begin{center}
\includegraphics[width=1\textwidth]{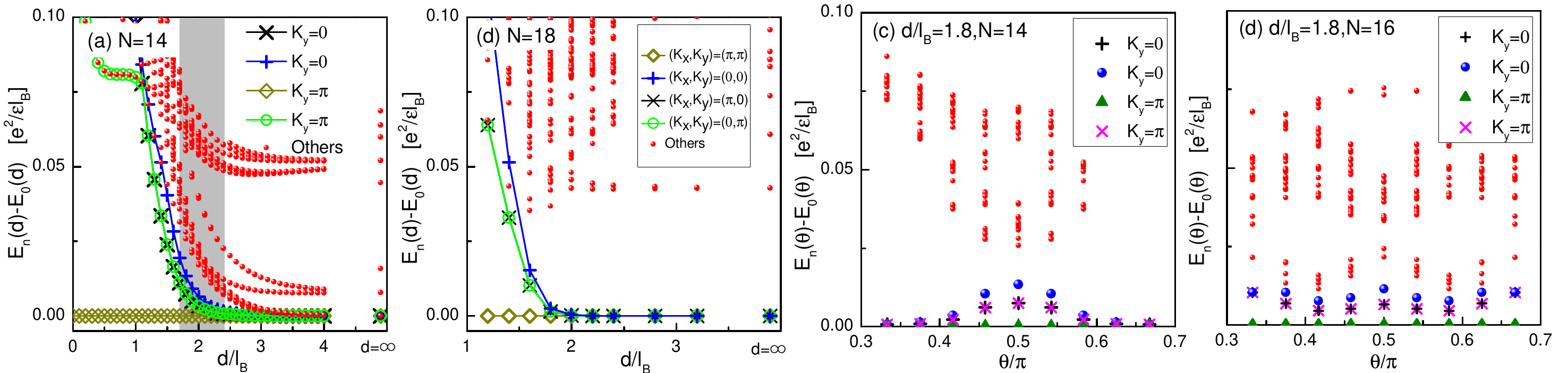}
\end{center}
\par
\renewcommand{\figurename}{Fig.}
\caption{ (Color online)  The energy spectrum as a function of layer distance $d/l_B$ for  (a) $N=14$ and (b) $N=18$ systems. For comparison, the $N=\infty$ data are shown on the rightmost side of each figure. The light grey area indicates the region  with a small gap between the lowest four states and higher energy states.  For $d/l_B=1.8$ , (c) and (d) show the energy spectrum as a function of the aspect angle $\theta$  of the unit cell
on torus for $N=14$ (c) and $N=16$ (d) systems with layer distance $d/l_B=1.8$. }
\label{Fig:E-d}
\end{figure*}
\begin{figure}[htbp]
\begin{center}
\includegraphics[width=0.6\textwidth]{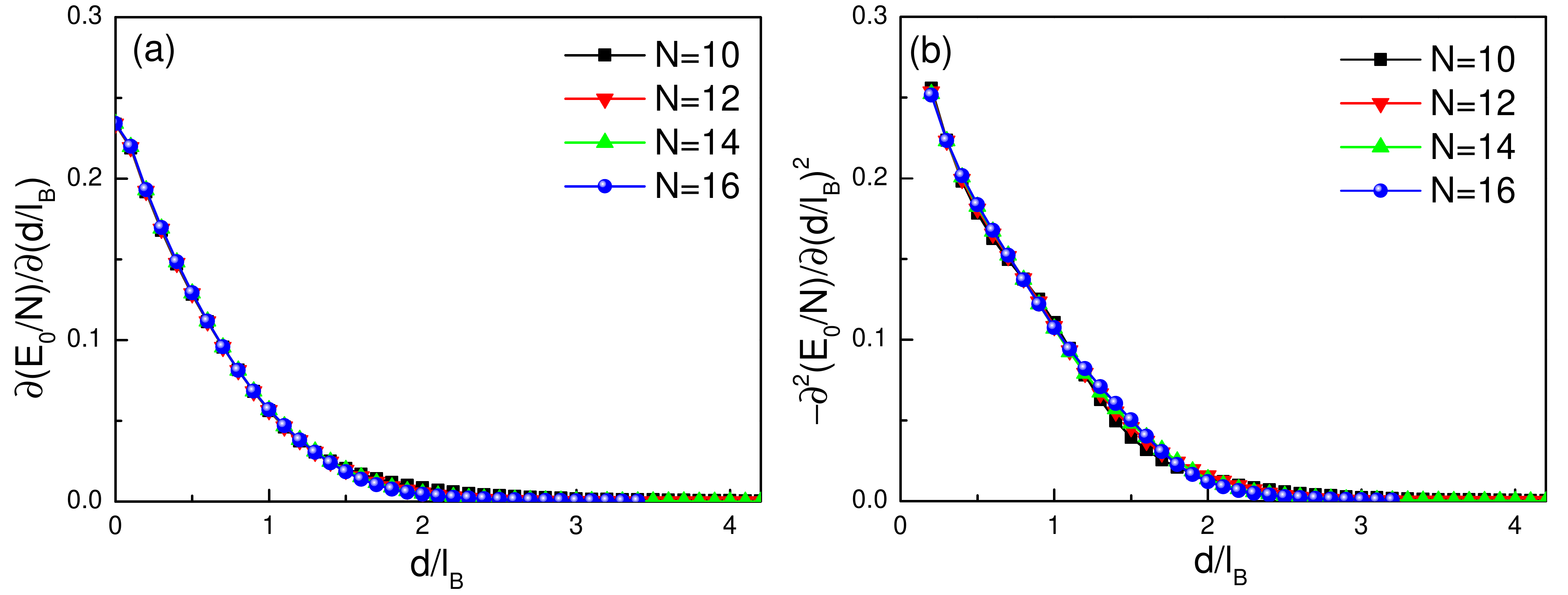}
\end{center}
\par
\renewcommand{\figurename}{Fig.}
\caption{(Color online) The first-order (a) and second-order (b) derivative curves of  ground-state energy $E_0/N$ as a function of  layer distance $d/l_B$. The smooth curves indicate the quantum phase transitions identified in this work  may be higher order continuous transitions.}
\label{Fig5:E0}
\end{figure}

\section{Discussion the finite size effect}
In this section,  we address the finite size effect for larger distance in CFL state. From the low energy spectrum shown in  Fig.~\ref{Fig:E-d} (a), we find indications of  the possible four-fold degenerate states in the regime $d/l_B\approx 1.8\sim 2.4$.
We also calculate the energy spectrum at infinite distance [see the rightmost data points in Fig.~\ref{Fig:E-d} (a)] to see the evolution of the energy spectrum to the decoupled limit. The four fold degeneracy  is generally  present  for two decoupled CFLs  due to the center of mass degeneracy of the electrons separating with other excited states
by a gap for finite-size systems. We further calculate the spectrum for $N=18$ system [see Fig.~\ref{Fig:E-d} (b)] , which has  $9$ electrons in each layer forming a completely filled shell in the  $3\times3$  momentum space besides the center of mass double degeneracy in each layer,
leading to an extra large gap between the lowest four states and excited states.  This indicates the finite size effect  and the four-fold degenerate states at finite  $d>d_{c_2}$ are  smoothly connected to the states
at the decoupled limit. The other method to check the robustness of the degeneracy and the finite excitation gap is to change the shape of the unit cell. We obtain the low energy spectrum  by varying the aspect angle of the unit cell on torus from
$\theta=\pi/3$ to $2\pi/3$, as shown in Fig.~\ref{Fig:E-d} (c) and (d) for $N=14$ and $N=16$ systems, respectively. The fourfold degeneracy does not change with tuning the geometry of unit cell from hexagon to square for $N=14$ while it disappears for $N=16$ system. The $N=14$ with hexagon unit cell is similar to  $N=18$ system with square unit cell with complete shell filling in momentum space.
Based on these analysis, we conjecture that  the four-fold degeneracy is due to strong finite-size effect, and  the vanishing superfluidity and drag Hall conductance indicate that  the quantum state on the $d>d_{c_2}$ side is already in the CFL phase.

Another open issue which is related to finite size effect is the transition type in $\nu=1/2+1/2$ bilayer quantum Hall systems. We also comment that the three phases we identified here are very robust based on various measurements. However, although we approach the system size up to 20 electrons by ED, the ground state energy evolves smoothly with layer distance for both the first and second order derivatives, as shown in Fig.~\ref{Fig5:E0}.   Thus the transition between different phases can be higher ordered continuous transitions, possibly of the Kosterlitz-Thouless type, which we leave for future study.

\end{document}